\documentclass[journal,twoside,web]{ieeecolor2}
\usepackage{generic}
\usepackage{cite}
\usepackage{amsmath,amssymb,amsfonts}
\usepackage{algorithmic}
\usepackage{graphicx}
\usepackage{textcomp}
\usepackage{array}
\usepackage{makecell}
\usepackage[table]{xcolor}
\usepackage[T1]{fontenc}
\def\BibTeX{{\rm B\kern-.05em{\sc i\kern-.025em b}\kern-.08em
    T\kern-.1667em\lower.7ex\hbox{E}\kern-.125emX}}
\begin{document}
\title{A Portable Multichannel Kilohertz Current Stimulator for Selective Peripheral Transcutaneous Stimulation}
\author{Jiatong Jiang, \IEEEmembership{Student Member, IEEE}, Andrea Bottin, Simos Koutsoftidis, \IEEEmembership{Student Member, IEEE}, Enrico Merlo, Georgios Gryparis, Yongkun Zhao,  \IEEEmembership{Student Member, IEEE},
Deren Y. Barsakcioglu, \IEEEmembership{Member, IEEE}, Dario Farina, \IEEEmembership{Fellow, IEEE}, and E. M. Drakakis, \IEEEmembership{Member, IEEE}
\thanks{This work was supported in part by a departmental Ph.D. 
scholarship from the Department of Bioengineering, Imperial College London, and in part by Imperial College Confidence in Concept under Grant PA1605.}
\thanks{\textit{(Deren Y. Barsakcioglu, Dario Farina, and E. M. Drakakis are co-senior authors.) (Corresponding authors: Dario Farina and E. M. Drakakis.)(Email: d.farina@imperial.ac.uk, e.drakakis@imperial.ac.uk.)} }
\thanks{Jiatong Jiang, Simos Koutsoftidis, Georgios Gryparis, Dario Farina, and E. M. Drakakis are with Department of Bioengineering, Imperial College London, London SW7 2AZ, UK. Andrea Bottin and Enrico Merlo are with OT Bioelettronica, Torino 10134, Italy. Yongkun Zhao is with the Centre for Machine Vision and Signal Analysis, University of Oulu, 90570 Oulu, Finland. Deren Y. Barsakcioglu is with the Department of Engineering Technology, Technical University of Denmark, 2750 Ballerup, Denmark.}}

\maketitle

\begin{abstract}
We present SineStim, a portable, 12-channel current stimulator for transcutaneous spinal cord stimulation and peripheral electrical stimulation at kilohertz frequencies. Each channel delivers independent, current-controlled sinusoidal waveforms with amplitudes from 0 to 50 mA (0.1 mA resolution) and frequencies from 0 to 50 kHz (0.1 Hz resolution), with burst modulation modes supported. A custom output stage with a high compliance voltage of $\pm$ 120 V was designed and developed, with galvanically isolated channels and independently programmable stimulation parameters for each channel. The stimulator performance was tested on both passive loads and human subjects. Benchtop characterisation on resistive and resistive-capacitive loads demonstrated a total harmonic distortion between 1 - 8 \% across typical operating conditions. Multichannel functionality was demonstrated in two-channel human forearm stimulation experiments. Burst-modulated waveforms with differing channel amplitudes modulated inter-finger force ratios, and channels with small frequency offsets elicited temporal interference force patterns at the beat frequency. Smoothly enveloped bursts of kilohertz sinusoidal waveforms produced negligible stimulation artifacts at steady state in concurrent surface electromyography (sEMG) recordings at motor threshold, in contrast to conventional biphasic square wave stimulation. SineStim delivers precise, isolated, multichannel kilohertz stimulation through a portable device, with performance demonstrated on both benchtop loads and human participants. By combining multichannel spatial control with minimal-artifact sEMG compatibility, SineStim enables future precision non-invasive neural stimulation paradigms based on multichannel optimisation and real-time closed-loop control capabilities not available via existing single-channel stimulators.
\end{abstract}


\section{Introduction}
\label{sec:introduction}
\IEEEPARstart{E}{lectrical} stimulation of the nervous system is widely applied to treat neurological conditions, including Parkinson’s disease, essential tremor, chronic pain, spinal cord injury, and stroke \cite{davidson2024neuromodulation, sdrulla2018spinal, marquez2020functional}. Stimulation can be delivered at the brain, spinal cord, or peripheral level. Invasive stimulation methods such as deep brain stimulation, epidural electrical stimulation and peripheral nerve stimulation with implanted electrodes can target neural structures with high spatial precision, but the surgical risk and the invasive nature of these procedures limit wider clinical and research access \cite{lozano2019deep}. Non-invasive surface techniques such as transcranial direct or alternating current stimulation, transcutaneous spinal cord stimulation (tSCS), and peripheral electrical stimulation (PES) offer safer alternatives, but at the cost of substantially reduced spatial selectivity \cite{woods2016technical}.

The full potential of non-invasive electrical stimulation has yet to be realised, particularly for spinal and peripheral applications. Firstly, although conventional stimulation is commonly delivered through a single channel, multichannel configurations can support spatially distributed asynchronous stimulation to reduce muscle fatigue, concurrent control of multiple neural pathways or spinal segments, and improved spatial selectivity \cite{maneski2013surface,shin2019multichannel,gerasimenko2015initiation,sharma2023multi}. A high channel count provides additional spatial degrees of freedom for model-based optimisation of multielectrode stimulation patterns, potentially improving neural recruitment selectivity. Secondly, kilohertz-frequency waveforms exhibit properties distinct from conventional carrier-free square pulses. The high stimulation intensities required for tSCS can produce substantial cutaneous sensation and activation of neural and neuromuscular structures beneath the surface electrodes. Kilohertz-frequency waveforms encounter reduced skin impedance and may produce less-attenuated electrical potentials in deeper tissues, providing a rationale for their use in transcutaneous stimulation of deeper neural structures \cite{medina2014volume}. Compared with conventional pulses, kilohertz-frequency waveforms have also been shown to bias recruitment towards motor efferent relative to proprioceptive afferent fibres \cite{keesey2026fundamental}. Furthermore, kilohertz-frequency carriers also form the basis of temporal interference (TI) stimulation \cite{grossman2017noninvasive}, although recent findings show controversies over the effectiveness of peripheral TI \cite{opanvcar2025same}. Lastly, kilohertz sinusoidal carriers combined with smooth burst envelopes may retain efficacy in eliciting neural responses while reducing recorded stimulation artifact by concentrating stimulation energy above the electromyography (EMG) bandwidth, consistent with findings from cardiac cells stimulation studies\cite{dura2012high}. This may facilitate concurrent low-artifact EMG recording and real-time feedback control with minimal loss of physiological signals. Collectively, these capabilities motivate the development of a multichannel stimulator with kilohertz sinusoidal carriers for PES and tSCS.

Several existing systems provide subsets of these capabilities. Multichannel, high-frequency brain stimulation platforms have emerged in recent years \cite{zhang2022designing} \cite{yan2026development}, driven in part by the growing research interest in multichannel TI stimulation \cite{grossman2017noninvasive} \cite{mirzakhalili2020biophysics} \cite{song2021multi} \cite{botzanowski2025focal}. However, brain stimulators cannot be directly repurposed for peripheral applications. Non-invasive brain stimulation with TI typically operates at subthreshold intensities of less than 4 mA \cite{demchenko2025human}, whereas PES and tSCS often requires suprathreshold current ranging from tens of mA to 100 mA in order to penetrate the skin and the subcutaneous tissue. Therefore, a much higher voltage compliance is needed \cite{martin2021utility}. Common bench stimulators used in PES and tSCS studies remain single-channel desktop systems, including the DS5 and DS8R (Digitimer Ltd., Welwyn Garden City, U.K.) \cite{kharboush2026transcutaneous,atkinson2022characterization,botzanowski2022noninvasive}. Custom multiplexing systems have been combined with single-channel biphasic stimulators to route pulses across multiple electrodes \cite{de2026spinal,powell2023epidural}. However, these systems cannot generate simultaneous continuous kilohertz waveforms with independently controlled carrier frequencies, phases, amplitudes, and burst envelopes. Multichannel stimulators delivering biphasic square pulses have also been reported \cite{xu2011programmable, trout2023portable, sheeraz2026bidirectional, alam2025openxstim, grishin2017five}, among which a few offer bursts of kilohertz square waves \cite{alam2025openxstim, grishin2017five}. However, no reported platform combines a high channel count with independently programmable, sinusoidal carriers over a wide kilohertz frequency range at hundreds of volts compliance. A high channel-count stimulator could enable spatial selectivity through modelling and optimisation. In this work, we address this gap by developing and presenting such a stimulation platform, thus enabling the design of spatially and temporally coordinated high-frequency stimulation and supporting the development of novel closed-loop transcutaneous stimulation strategies.

\begin{figure*}[!t]
\includegraphics[width=\linewidth]{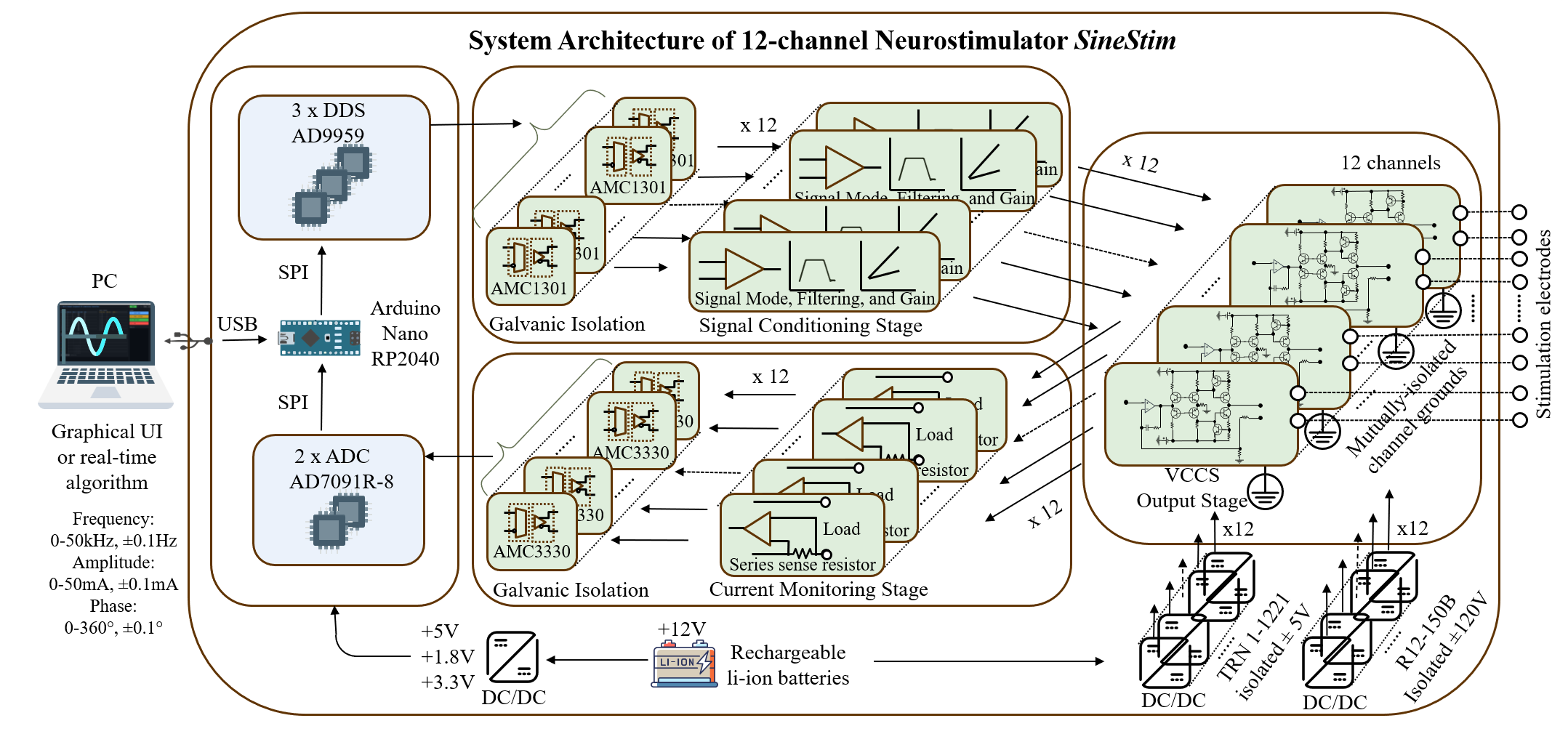}
\caption{System-level block diagram of SineStim. A PC communicates with an on-board microcontroller (Arduino Nano RP2040) via USB, which programs three four-channel DDS chips (AD9959) over SPI. Each of the 12 channels benefits from a galvanically isolated signal conditioning path, a voltage-controlled current source (VCCS) output stage with ±120 V isolated supply (R12-150B), and a current monitoring path feeding back to two eight-channel ADCs (AD7091R-8). All channel output stages have mutually isolated grounds. The system is powered by a rechargeable 12 V lithium-ion battery.}
\label{fig1}
\end{figure*}

In this paper, we present SineStim, a portable, programmable, high-compliance, 12-channel kilohertz-frequency current-controlled stimulator for non-invasive neural stimulation. Each channel delivers sinusoidal waveforms up to 50 mA in amplitude and 50 kHz in frequency. Instead of multiplexing, each channel amplitude, frequency, phase, and burst modes are independently programmable in real time. The paper is organised as follows. Section II describes the hardware architecture and the analytical model of the output stage. Section III presents the characterisation results on resistive and resistive-capacitive loads representative of the electrode-skin interface, as well as validation of stimulation response in human participants. Section IV discusses the experimental results while Section V concludes the paper.

\section{Methods}

\subsection{Hardware Architecture}
The system-level block diagram is shown in Fig.~\ref{fig1}. Three AD9959 four-channel direct digital synthesis (DDS) chips (Analog Devices, Wilmington, MA), each with 32-bit frequency tuning, 14-bit phase offset, and 10-bit amplitude scaling registers, generate precision sinusoidal waveforms for 12 independent channels. These specifications yield a frequency resolution of 0.1 Hz and an amplitude resolution of at least 0.1 mA at the output. The differential DDS outputs pass through signal conditioning stages for single-ended conversion, bandpass filtering, and gain adjustment. The conditioned signal serves as the input to the output stage, a voltage-to-current converter (VCCS) with ±120 V compliance, described in detail in Section II-B.
A $0.5~\Omega$ sense resistor is placed in series with each electrode load. The voltage across this resistor is amplified by an INA241A1 current-sense amplifier (Texas Instruments, Dallas, TX), and the outputs from all 12 channels are digitised by two AD7091R-8 eight-channel ADCs (Analog Devices) for real-time current monitoring. Both the ADC readback and DDS programming are managed by an onboard Arduino Nano RP2040 microcontroller (Arduino, Monza, Italy) via SPI, which in turn communicates with a host PC over a USB serial interface. The Arduino also controls isolated relay modules that switch the ±120 V rail of each channel. These relays can also be activated via a physical emergency stop button. The device is powered by an 11.1 V, 5.2 Ah, six-cell rechargeable lithium-ion battery pack (RS PRO, RS Components, Corby, UK), enabling fully portable operation. Power sequencing of DC/DC modules is employed to reduce inrush current. The battery pack provides maximum 5 hours of operation under typical experimental conditions, with battery life varying depending on the number of active channels and stimulation amplitude.
As indicated in Fig.~\ref{fig1}, the high-voltage output stages are galvanically isolated from all upstream and downstream signal paths by means of AMC3330 and AMC1301 isolation amplifiers (Texas Instruments). Mutual isolation between channel output stages is achieved through dedicated, per-channel isolated power supplies: R12-150B modules (RECOM Power, Gmunden, Austria) provide the ±120 V rails, and TRN 1-1221 modules (TRACO Power, Baar, Switzerland) provide ±5 V. Each channel has a floating ground implemented on a separate ground plane zone on the PCB.  

\begin{figure}[!t]
\centerline{\includegraphics[width=\columnwidth]{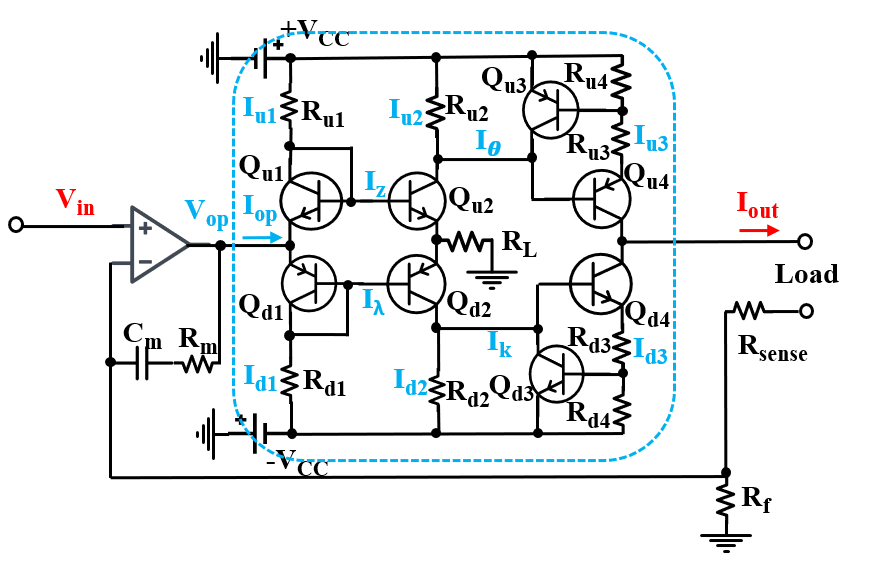}}
\caption{Voltage-to-current converter (VCCS) based output stage topology of one stimulation channel. The quantities $I_{op}$, $I_{u1}$, $I_{d1}$, $I_{z}$, $I_{\lambda}$, $I_{u2}$, $I_{d2}$, $I_{\theta}$, $I_{k}$, $I_{u3}$, $I_{d3}$, $I_{out}$ denote currents of corresponding branches, and $V_{in}$ and $V_{op}$ denote voltages of corresponding nodes. $V_{cc}$ is 120 V.}
\label{fig2}
\end{figure}

\begin{table}
\centering
\caption{Component values}
\label{table}
\setlength{\tabcolsep}{12pt}
\begin{tabular}{|p{40pt}|p{65pt}|p{40pt}|}
\hline
Component& 
Symbol / Parameter& 
Value \\
\hline
$R_{u1}$, $R_{d1}$& 
$R_{1}$& 
$10~M\Omega$  \\
$R_{u2}$, $R_{d2}$& 
$R_{2}$& 
$220~k\Omega$  \\
$R_{u3}$, $R_{d3}$& 
$R_{3}$& 
$150~\Omega$  \\
$R_{u4}$, $R_{d4}$& 
$R_{4}$& 
$3~\Omega$  \\
$R_{L}$&$R_{L}$
& 
$1~k\Omega$  \\
$R_{sense}$&$R_{sense}$ 
& 
$0.5~\Omega$  \\
$R_{f}$ & $R_{f}$
& 
$6~\Omega$  \\
$R_{m}$ & $R_{m}$
& 
$50~\Omega$  \\
$C_{m}$&$C_{m}$ 
& 
$3~nF$  \\
$Q_{u4}, Q_{d4}$ & $\beta_{u4}, \beta_{d4}$
& 
$30-240$  \\
\hline
\end{tabular}
\label{values}
\end{table}

\subsection{Voltage-Controlled Current Source Output Stage}
This section presents a practical design and analysis framework for a high-compliance, kilohertz-range, low-distortion VCCS output stage with current limitation. As shown in Fig.~\ref{fig2}, the circuit consists of a high-voltage complementary Class-AB bipolar junction transistor (BJT) output stage (in blue box), enclosed within an op-amp current-feedback loop for voltage-to-current conversion with Miller compensation. The high-voltage BJT stage determines output compliance, current-drive capability, open-loop distortion, and stability margin, whereas $1/R_f$ sets the closed-loop transconductance:  
\begin{IEEEeqnarray}{rCl}
I_{\mathrm{out}} & \approx & \frac{V_{\mathrm{in}}}{R_f}. \label{eq:iout_simple}
\end{IEEEeqnarray}
Here, $V_{in}$ is the command voltage, $V_{op}$ is the op-amp output voltage driving the BJT stage, and $I_{out}$ is the delivered stimulation current.
Resistor values for the upper and lower branches are implemented symmetrically. Relevant component values and parameters are detailed in Table.~\ref{values}.

\subsubsection{Class-AB Translinear Biasing}
The Class-AB pre-driver comprises the complementary translinear loop formed by discrete power BJTs $Q_{u1}$, $Q_{d1}$, $Q_{u2}$, and $Q_{d2}$. Assuming forward-active operation, common temperature, and sufficiently large current gain, along with approximate matching for the NPN and PNP transistor pairs $Q_{u1}/Q_{u2}$ and $Q_{d1}/Q_{d2}$ , the following relationships can be derived for the collector currents of $Q_{u2}$ and $Q_{d2}$:
\begin{IEEEeqnarray}{rCl}
I_{C,Q_{u2}} & = & \frac{\dfrac{V_{op}}{R_{L}} + \sqrt{\left(\dfrac{V_{op}}{R_{L}}\right)^{2} + 4\left(\dfrac{V_{CC}}{R_1}\right)^{2}}}{2}, \label{eq:icqu2} \\
I_{C,Q_{d2}} & = & \frac{-\dfrac{V_{op}}{R_{L}} + \sqrt{\left(\dfrac{V_{op}}{R_{L}}\right)^{2} + 4\left(\dfrac{V_{CC}}{R_1}\right)^{2}}}{2}. \label{eq:icqd2}
\end{IEEEeqnarray}
Note that 
\begin{IEEEeqnarray}{rCl}
I_{C,Q_{u2}} - I_{C,Q_{d2}} = \frac{V_{op}}{R_{L}}.\label{eq:icqu12} 
\end{IEEEeqnarray}

\subsubsection{Two Limiting Output-Stage Regimes}
The next stage within the open-loop amplifier converts the pre-driver currents $I_{C,Qu2}$ and $I_{C,Qd2}$ into the output current $I_{out}$. To calculate $I_{out}$ and therefore the open-loop gain, the relationships between intermediate branches $I_{u2,d2}$, $I_{\theta, k}$, and $I_{u3,d3}$ need to be determined first. Two limiting cases should be considered.

\textit{Case A: Resistor-dominated Regime.}
Considering the case when
\begin{IEEEeqnarray}{c}
I_\theta \ll I_{u2}, \qquad I_k \ll I_{d2} \label{eq:caseA_cond}
\end{IEEEeqnarray}
It can be calculated that the value of the open-loop transconductance $G_{A}$ is mainly determined by local resistor ratios:
\begin{IEEEeqnarray}{rCl}
G_A = \frac{I_{\mathrm{out}}}{V_{op}} & \approx & \frac{R_2}{(R_3 + R_4)\,R_{L}}. \label{eq:GA}
\end{IEEEeqnarray}

\textit{Case B: Current-gain-dominated Regime.} In the opposite case, when
\begin{IEEEeqnarray}{c}
I_\theta \gg I_{u2}, \qquad I_k \gg I_{d2} \label{eq:caseB_cond}
\end{IEEEeqnarray}
the open-loop transconductance $G_{B}$ depends strongly on the current gains of $Q_{u4}$ and $Q_{d4}$. It can be shown that:
\begin{IEEEeqnarray}{rCl}
G_B = \frac{I_{\mathrm{out}}}{V_{op}} & \approx & \frac{\beta_{u4,d4}}{R_{L}}. \label{eq:GB}
\end{IEEEeqnarray}
Note that the effect of any mismatch between the gains of positive and negative phases $\beta_{u4}$ and $\beta_{d4}$ in this case could be corrected by the subsequent closed-loop feedback.

\begin{figure*}[!t]
\centerline{\includegraphics[width=\linewidth]{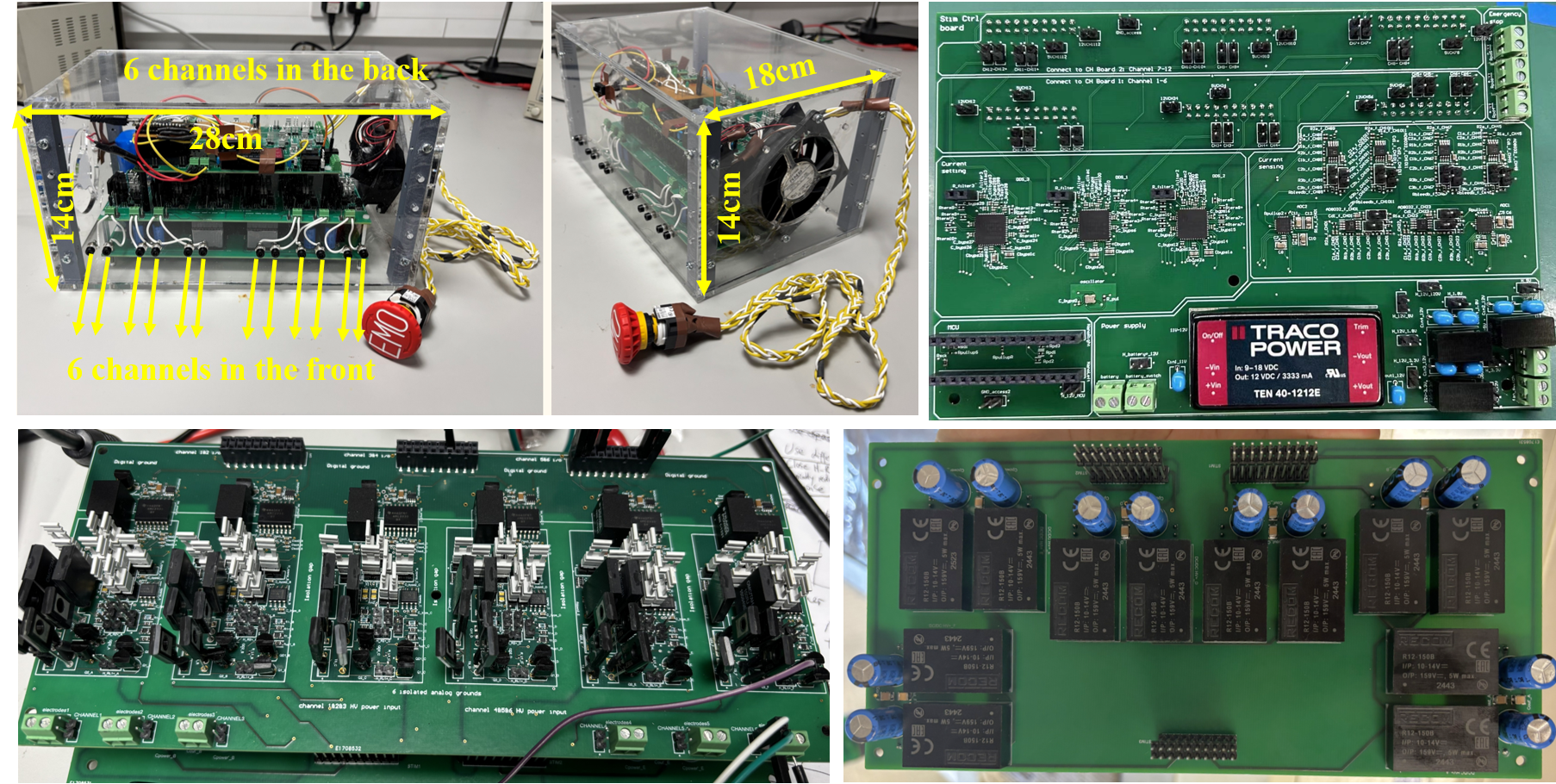}}
\caption{Fabricated SineStim device and individual printed circuit boards. Top left and top middle: Front and side views of the assembled device within an acrylic enclosure, containing a three-PCB stack. Top right: Digital control and low-level power management PCB (top of stack), interfacing with two output stage PCBs to drive all 12 channels. Bottom left: six-channel output stage PCB containing analogue VCCS and current monitoring (middle of stack). Bottom right: Six-channel isolated $\pm$120V PCB (bottom of stack).}
\label{fig3}
\end{figure*}

\subsubsection{Hardware Current Limitation}
Protection transistors $Q_{u3}$ and $Q_{d3}$ provide hardware current limiting via local negative feedback. When the voltage across $R_{u4}$ or $R_{d4}$ approaches the base-emitter turn-on voltage, $Q_{u3}$ and $Q_{d3}$ start to conduct, and the base currents of $Q_{u4}$ and $Q_{d4}$ are reduced leading in turn to a reduction of the corresponding collector currents. The approximate current limits are
\begin{IEEEeqnarray}{rClCrCl}
I_{\mathrm{lim},u3} & \approx & \frac{V_{EB,Q_{u3}}}{R_{4}}, & \quad & I_{\mathrm{lim},d3} & \approx & \frac{V_{BE,Q_{d3}}}{R_{4}}. \label{eq:current_limit}
\end{IEEEeqnarray}
\subsubsection{Closed-Loop Transfer Function}

Equations \eqref{eq:GA} and \eqref{eq:GB} give the open-loop transconductances \(G_A\) and \(G_B\), respectively, from \(v_{\mathrm{op}}\) to \(i_{\mathrm{out}}\) under two limiting conditions. The loop is closed through the shunt-feedback op-amp pathway and stabilized by Miller compensation. The resulting first-order closed-loop transconductance is
\begin{IEEEeqnarray}{rCl}
\frac{I_{\mathrm{out}}(s)}{V_{\mathrm{in}}(s)}
&=&
\frac{R_m+R_f}
     {R_f\left(R_m-\frac{1}{G_i}\right)}
\nonumber\\
&& {}\times
\frac{
s+\left[C_m\left(R_m+R_f\right)\right]^{-1}
}{
s+\left[C_m\left(R_m-\frac{1}{G_i}\right)\right]^{-1}
},
i=A,B.
\label{eq:closed_loop_transconductance}
\end{IEEEeqnarray}

Under the first-order stability condition:
\begin{IEEEeqnarray}{c}
R_m > \frac{1}{G_i}, \qquad i = A,B.
\label{eq:stability_condition}
\end{IEEEeqnarray}
The transfer function \eqref{eq:closed_loop_transconductance} has one negative zero and one negative pole, whose frequencies are: 
\begin{IEEEeqnarray}{rCl}
f_{z}
&=&
\frac{1}
     {2\pi C_m(R_m+R_f)},
     \qquad i = A,B.
     \label{eq:zero}
\\
f_{p}
&=&
\frac{1}
     {2\pi C_m\left(R_m-\frac{1}{G_i}\right)},
\qquad i = A,B.
\label{eq:pole}
\end{IEEEeqnarray}

\subsubsection{Design Criteria Concluded}

Equations~\eqref{eq:iout_simple}, \eqref{eq:current_limit},
\eqref{eq:stability_condition}, \eqref{eq:zero} and
\eqref{eq:pole} define the component-level design space, with \(G_{A}\) and \(G_{B}\) given by \eqref{eq:GA} and \eqref{eq:GB}. \(R_{f}\) is first chosen to set the closed-loop DC gain according to~\eqref{eq:iout_simple}. \(R_{4}\) is selected to set the desired protective current limit according to \eqref{eq:current_limit}. Given the practical chosen values of \(R_2\), \(R_3\), \(R_4\), and \(R_L\), \(R_m\) needs to satisfy the stability condition in \eqref{eq:stability_condition}, using the minimum specified current gain of \(Q_{u4}\) and \(Q_{d4}\) in Table.~\ref{values}.

For the first-order model with \(f_z<f_p\), \(R_m\) and \(C_m\) are selected sequentially. Increasing \(R_m\) reduces the ratio \(f_p/f_z\), thereby limiting gain rise and phase lead. \(C_m\) is then decreased according to \eqref{eq:zero} and \eqref{eq:pole} to shift both corner frequencies upward, compensating for the downward shift introduced by \(R_m\) and maintaining approximately constant closed-loop transconductance over the intended \(0\)--\(50~\mathrm{kHz}\) operating band.

Because the practical response is also
influenced by transistor parasitic capacitances, finite current gain,
PCB parasitics, and component mismatch, the first-order analysis is
used only as a component-selection guideline. The stability and performance of the implemented circuit is subsequently experimentally verified using representative resistive
and resistive-capacitive loads.

\subsection{Device Fabrication}
The fabricated SineStim device, shown in Fig.~\ref{fig3}, comprises three stacked 4-layer double-sided printed circuit boards (PCBs) housed in a custom 28 × 18 × 15 cm acrylic enclosure. The device integrates a lithium-ion battery pack, a DC cooling fan for the power output transistors, and an emergency stop switch (visible in red) positioned to be accessible to the subject at all times. Twelve electrode connectors are mounted on the front panel and twelve on the rear. Communication with a host PC is established via a USB connector.

\subsection{Electrical Characterisation}

The device output was first characterised using one resistive load of $1~k\Omega$ over three current amplitudes and five stimulation frequencies spanning the minimum, representative, and maximum operating values. The voltage drop across a $1~k\Omega$ resistor was measured using two channels of a mixed-signal oscilloscope (MSO-X 2002A, Keysight Technologies, Santa Rosa, CA, USA), through which the output current was then reconstructed.

Time-domain waveforms were obtained for all 15 stimulation conditions and compared. Their frequency spectra were also computed using the Fast Fourier transform (FFT). Signal quality was quantified using both total harmonic distortion (THD) and signal-to-noise ratio (SNR). Both THD and SNR were evaluated over the available analysis bandwidth up to 200 kHz, where the noise was defined as the RMS residual after subtracting the fitted DC, fundamental, and harmonic components from the measured waveform.

To assess output consistency under more realistic loading conditions, the selected stimulation settings were further tested on five resistive and resistive-capacitive (RC) loads representing typical bipolar electrode-skin interface impedances.

\subsection{Proof-of-Concept Experiments in Human Participants}

\subsubsection{Participants and Ethics}
Three healthy human participants (two females, one male, 27 ± 1 years old) were recruited for the stimulation experiments. The experimental procedures were conducted in accordance with the Declaration of Helsinki and approved by the Imperial College Research Ethics Committee (reference number: 20IC5945).
In order to demonstrate the stimulator's multichannel performance, we applied two-channel stimulation to the subject's forearm and measured their finger force responses. We also applied stimulation to the subject's upper arm and measured bipolar surface electromyography responses from the forearm to characterise stimulation artifact.

\begin{figure*}[!t]
\centerline{\includegraphics[width=\linewidth]{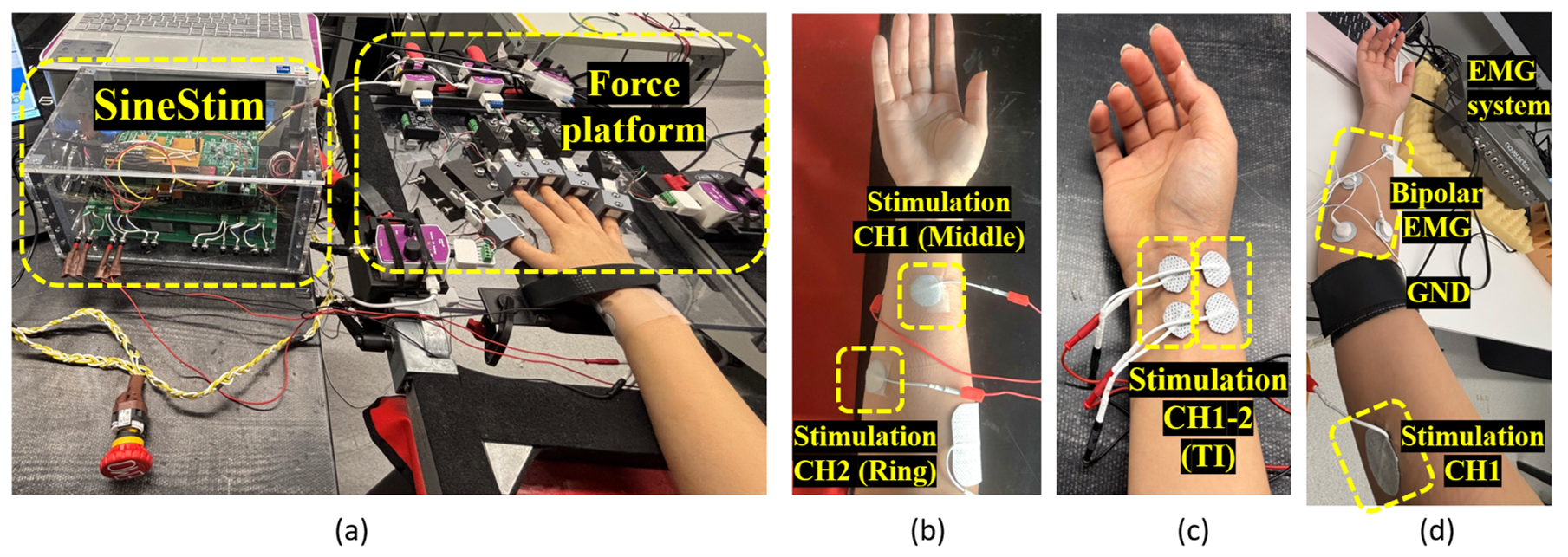}}
\caption{Experimental setups for three proof-of-concept human peripheral stimulation experiments. (a) Stimulation and finger force measurement setup. The subject's hand is placed on a custom force platform (Phlext), with the load cells in contact with both sides of fingers recording flexion and extension forces. Stimulation electrodes are connected to the PC-controlled SineStim device. (b) Electrode setup for two-channel motor-point stimulation for finger force ratio control. Highlighted are the active electrodes for channel 1 and channel 2, targeting middle- and ring-finger flexion motor points, respectively. Both channels share a 5 × 9 cm return electrode positioned over the extensor side of the forearm. (c) Electrode setup for two-channel temporal interference (TI) stimulation, comprising two parallel 1.5 cm diameter electrode pairs positioned near the wrist. Thumb flexion force is recorded during stimulation. (d) Stimulation and bipolar EMG recording setup for artifact characterisation experiment. Two bipolar EMG channels are placed over the forearm flexors, whereas bipolar stimulation is delivered to the medial distal upper arm through two self-adhesive hydrogel electrodes. The EMG ground is connected to the elbow band. Bipolar EMG signal is acquired via the Novecento system at 2000 samples/s.}
\label{fig4}
\end{figure*}

\subsubsection{Two-Channel Motor-Point Stimulation for Finger-Force Ratio Control}
\label{sec:ratio}

To demonstrate independent control of the current amplitude across channels, two-channel motor-point stimulation was used to modulate the relative flexion forces of the middle and ring fingers. The experimental setup is shown in Fig.~\ref{fig4} (a). Each subject was seated comfortably in a chair with their arm rest on the table. Each subject was asked to place their right hand onto the custom finger force platform Phlext. The load-cell force signals were amplified using Forza-B force amplifiers and recorded using the Novecento system (OT Bioelettronica, Torino, Italy). For each participant, the forearm motor points eliciting middle- and ring-finger flexion were identified. Two self-adhesive hydrogel electrodes (ValuTrode, Axelgaard Manufacturing Co., Ltd., Fallbrook, CA) with 3 cm diameter were placed on these identified motor points, as shown in Fig.~\ref{fig4} (b), while a common 5 cm x 9 cm electrode was placed on the extensor side of the forearm as the common return electrode for two channels. The motor threshold and maximum tolerable current were determined separately for each motor point and subject. Five equally spaced amplitude levels spanning this range were then defined for each channel. Across conditions A–E, the amplitude of Channel 1 decreased from its maximum tolerable current to its motor threshold, while that of Channel 2 increased from its motor threshold to its maximum tolerable current. Each condition was applied for 20 s, followed by a 50 s rest period, while the middle- and ring-finger flexion forces were recorded simultaneously.

\subsubsection{Two-Channel Temporal Interference Stimulation}

We then demonstrated the stimulator’s ability to independently set the carrier frequency of each channel using a two-channel TI paradigm. Two concurrent kilohertz-frequency currents separated by a small frequency offset produced an amplitude-modulated interference envelope at the difference frequency, which was used to evoke periodic motor response \cite{grossman2017noninvasive} \cite{opanvcar2025same}. The resultant finger force was recorded in real time during stimulation.

The setup is again shown in Fig.~\ref{fig4} (a). Four self-adhesive hydrogel electrodes with 1.5 cm diameter were placed near the wrist as shown in Fig.~\ref{fig4} (c) and connected to SineStim output Channels 1 and 2. Each subject was asked to place their right hand into the Phlext finger force platform and the force was recorded via Novecento, in the same way as Section \ref{sec:ratio}. Both channels delivered sinusoidal currents with amplitudes just above the subject's motor threshold. The frequency of Channel 1 was fixed at 10,000 Hz, whereas that of Channel 2 was set to 10,001, 10,002, or 10,005 Hz, generating target beat frequencies of 1, 2, and 5 Hz, respectively. The flexion force of the thumb was recorded simultaneously. The modulation related to the beat frequency in the measured force profile was expected to provide functional validation of the multichannel frequency control of the stimulator.

\subsubsection{Stimulation-Artifact Characterization With Smoothly Enveloped kHz Bursts}

Finally, concurrent bipolar EMG recording during smoothly enveloped kilohertz-frequency burst stimulation was evaluated. As shown in Fig.~\ref{fig4} (d), a bipolar electrode pair connected to one stimulation channel was placed on the upper arm near the elbow. Two bipolar EMG channels were placed over the forearm flexor area and connected to the Novecento system for recording, with the EMG ground connected to moistened elbow band. The channel with larger artifact was assessed and compared among three input waveforms. Hann-enveloped bursts of sinusoidal carriers, delivered at a burst repetition frequency of 80~Hz with a duration of 5~ms, were compared with conventional biphasic square pulses both with and without kilohertz square carriers delivered at the same repetition frequency. For each waveform, the stimulation amplitude was set to one third of their corresponding motor threshold. Stimulation was applied continuously for at least 10~s. The participant remained relaxed initially and subsequently performed voluntary wrist flexion while stimulation continued. The stimulation artifact in each case were assessed.

\section{Results}

\subsection{Output Current Waveform on Resistive and RC Load}

\begin{figure}[!t]
\centerline{\includegraphics[width=\columnwidth]{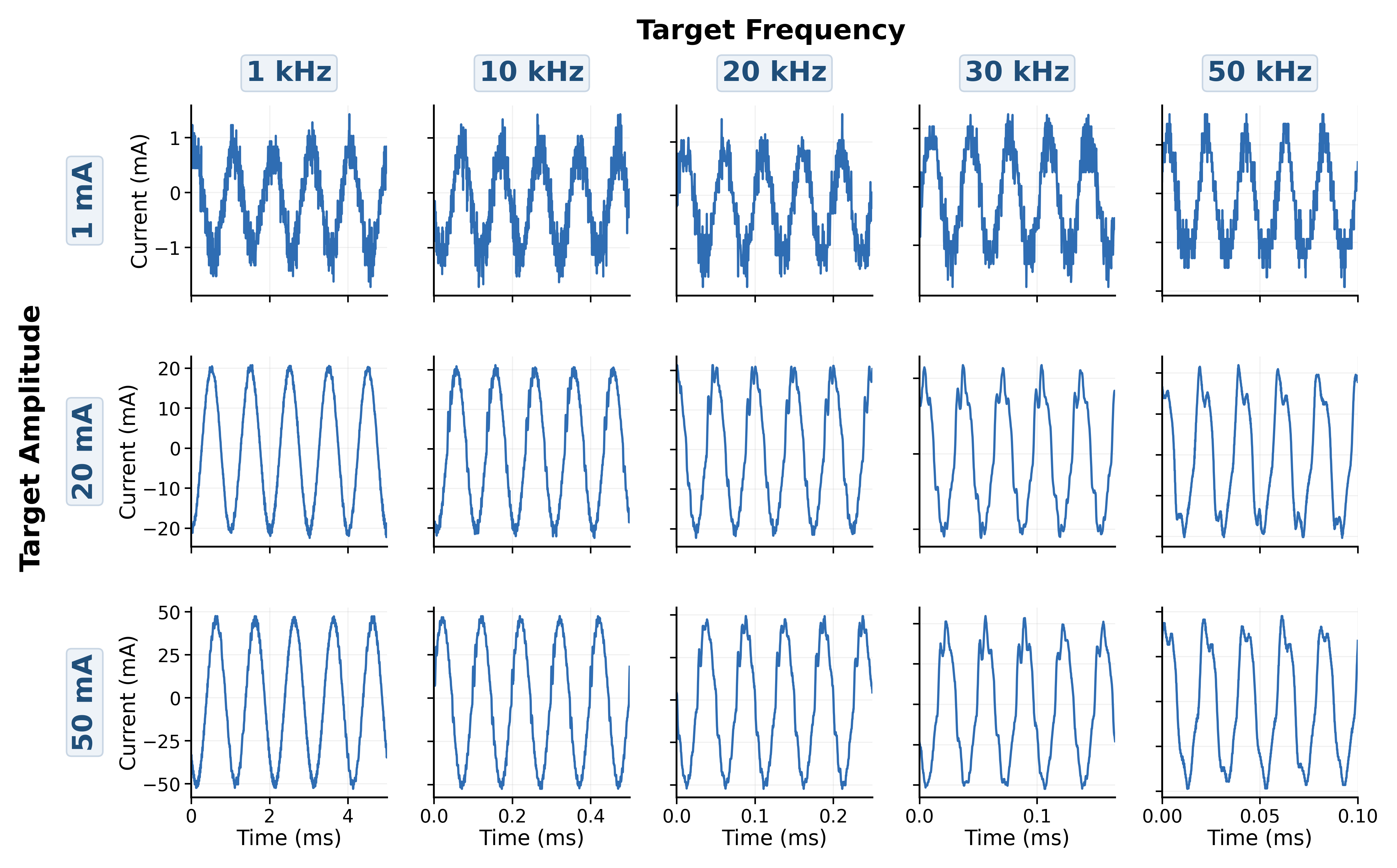}}
\caption{Measured output current waveforms on a $1~k\Omega$ resistive load across the full amplitude (1-50 mA) and frequency (1-50 kHz) operating range. All waveforms are normalised in both axes for visual comparison. The 1 mA row illustrates the noise floor limitation, whereas the 50 kHz column illustrates the harmonic distortion limitation at the upper frequency boundary.}
\label{fig5}
\end{figure}

\begin{figure}[!t]
\centerline{\includegraphics[width=\columnwidth]{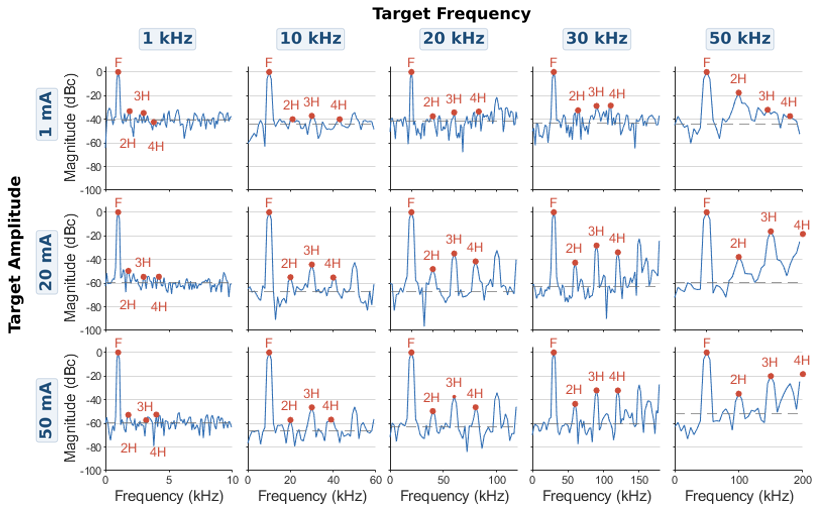}}
\caption{Frequency spectra (FFT) of the measured output current waveforms on a $1~k\Omega$ resistive load across the full amplitude (1-50 mA) and frequency (1-50 kHz) operating range. The fundamental (F) and first three harmonics (1H, 2H, and 3H) are marked in red. The approximate noise floor is indicated.}
\label{fig6}
\end{figure}

The time and frequency-domain output current waveforms on a $1~k\Omega$ resistive load across the full amplitude (1-50 mA) and frequency (1-50 kHz) operating range are shown in Fig.~\ref{fig5} and Fig.~\ref{fig6}, respectively. Five representative carrier frequencies were selected: 1 kHz, representing the lowest carrier frequency used experimentally; 10 kHz and 20 kHz, representing commonly used operating range; and 50 kHz, representing the upper carrier-frequency limit. Three representative amplitudes were also selected: 1 mA, representing the lowest output level; 20 mA, representing a typical amplitude for evoking motor responses; and 50 mA, representing the maximum output limit.

The time-domain plots in Fig.~\ref{fig5} show that when operating well within the compliance-voltage limit, the generated current waveform remains approximately sinusoidal under most tested conditions. The 1 mA row illustrates the greatest visible influence of the residual noise floor, whereas distortion becomes increasingly pronounced with increasing carrier frequency, particularly toward 50 kHz. The most well-preserved waveforms are observed in the lower-frequency, moderate-amplitude region of the operating space, including the commonly adopted conditions around 1-20 kHz and 10-20 mA.

\begin{figure}[!t]
\centerline{\includegraphics[width=\columnwidth]{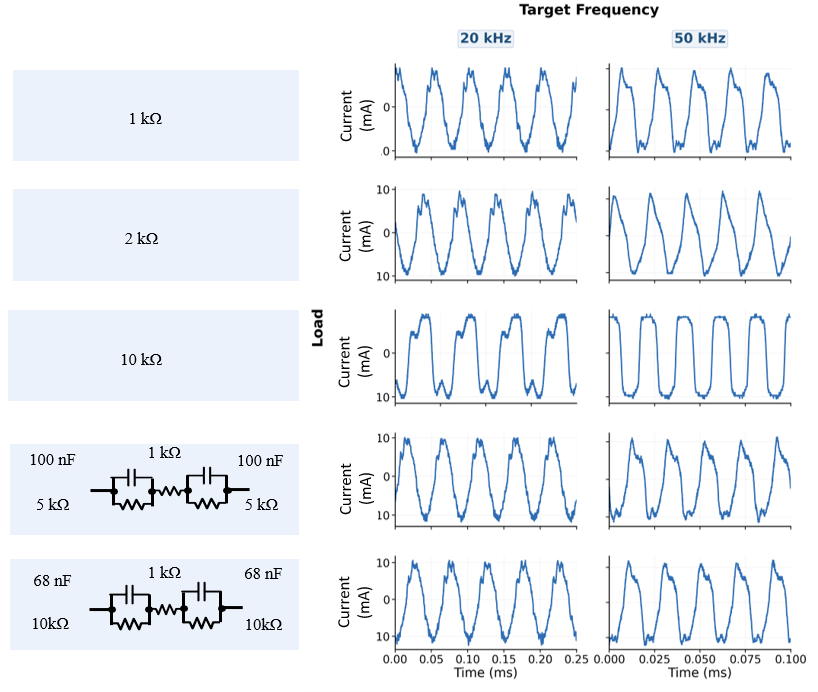}}
\caption{Measured output current waveforms at 10 mA for 20 kHz (typical) and 50 kHz (maximum) into resistive and resistive-capacitive (RC) loads. Resistive loads range from $1~k\Omega$ to $10~k\Omega$; Waveform fidelity is generally maintained across all loads, except at $10~k\Omega$, where the required output voltage nearly approaches the ±120 V compliance limit, producing visible near-saturation distortion.}
\label{fig7}
\end{figure}

\begin{table}[!t]
\definecolor{lightblue}{RGB}{204, 229, 255}
\caption{Measured THD and SNR (< 200 kHz) of the output current on a $1~k\Omega$ resistive load across the full amplitude (1-50 mA) and frequency (1-50 kHz) operating range. Highlighted in blue are the most common operating conditions (20-50 mA and 1-20 kHz).THD remains between 1-8 \% in highlited conditions.}
\label{tab:thd_snr}
\centering
\footnotesize
\setlength{\tabcolsep}{7pt}
\renewcommand{\arraystretch}{3}

\begin{tabular}{|c|c|c|c|c|c|c|}
\hline
\shortstack{\textbf{Target}\\\textbf{amplitude}} &
\textbf{Metric} &
\textbf{1 kHz} &
\shortstack{\textbf{10}\\\textbf{kHz}} &
\shortstack{\textbf{20}\\\textbf{kHz}} &
\shortstack{\textbf{30}\\\textbf{kHz}} &
\shortstack{\textbf{50}\\\textbf{kHz}} \\
\hline

\raisebox{-1.5ex}{\textbf{1 mA}} &
\shortstack{\textbf{THD}\\\textbf{(\%)}} &
12.9 & 6.6 & 7.1 & 4.8 & 14.7 \\
\cline{2-7}
&
\shortstack{\textbf{SNR}\\\textbf{(dB)}} &
10.8$^{*}$ & 18.5 & 17.9 & 19.9 & 20.6 \\
\hline

\raisebox{-1.5ex}{\textbf{20 mA}} &
\shortstack{\textbf{THD}\\\textbf{(\%)}} &
\cellcolor{lightblue}1.2 & \cellcolor{lightblue}6.1 & \cellcolor{lightblue}8.3 & 13.9 & 15.4 \\
\cline{2-7}
&
\shortstack{\textbf{SNR}\\\textbf{(dB)}} &
\cellcolor{lightblue}28.3$^{*}$ & \cellcolor{lightblue}40 & \cellcolor{lightblue}38.1 & 34.2 & 30.4 \\
\hline

\raisebox{-1.5ex}{\textbf{50 mA}} &
\shortstack{\textbf{THD}\\\textbf{(\%)}} &
\cellcolor{lightblue}1.5 & \cellcolor{lightblue}4.4 & \cellcolor{lightblue}6.6 & 10.2 & 9.9 \\
\cline{2-7}
&
\shortstack{\textbf{SNR}\\\textbf{(dB)}} &
\cellcolor{lightblue}28.2$^{*}$ & \cellcolor{lightblue}34.6 & \cellcolor{lightblue}33 & 25.7 & 25.2 \\
\hline

\end{tabular}

\vspace{3pt}
\parbox{\columnwidth}{\footnotesize
$^{*}$The 1-kHz SNR values were evaluated over the available
100-kHz bandwidth because of the acquisition sampling rate.}

\end{table}

\begin{figure*}[!t]
\centerline{\includegraphics[width=\linewidth]{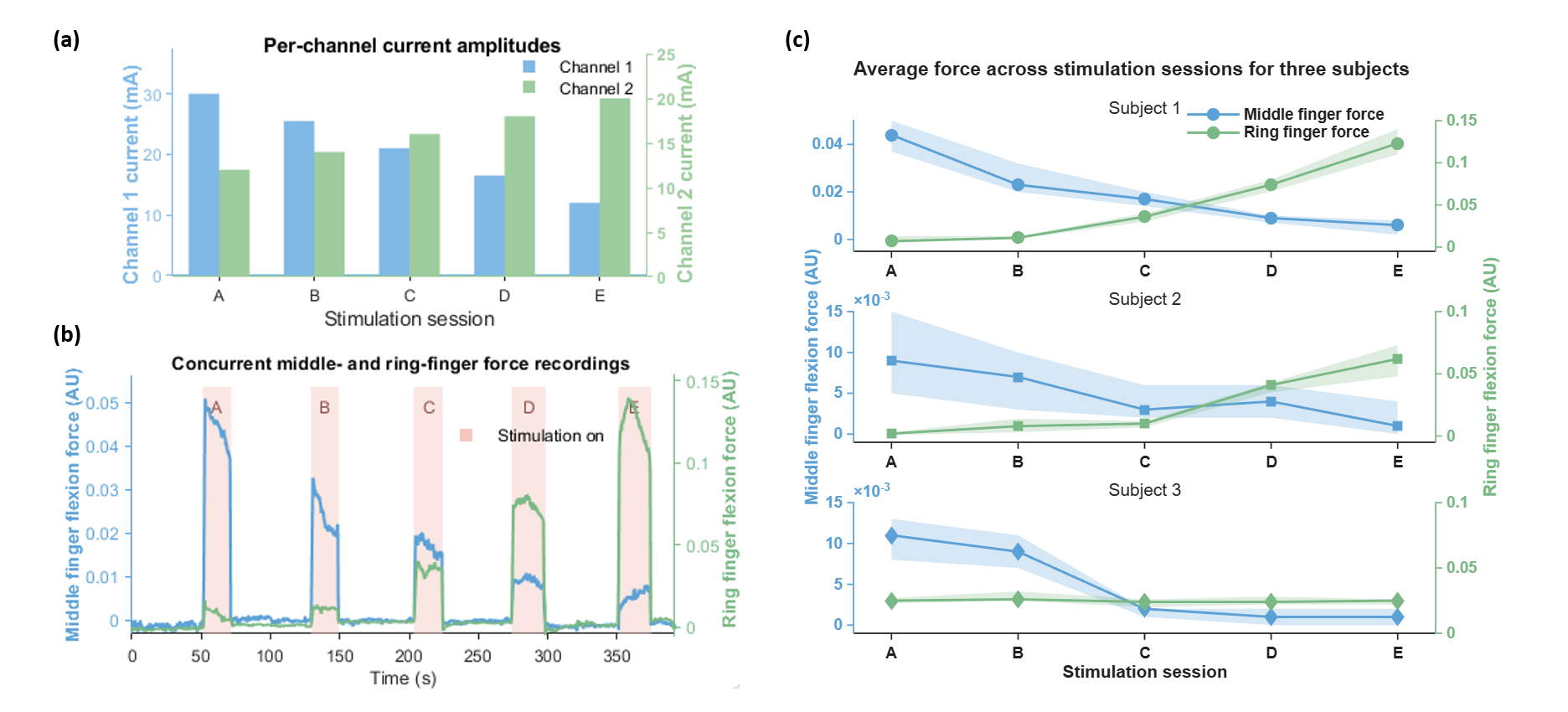}}
\caption{Multichannel stimulation enables fine control of the relative contributions of individual movements. Composite motor-pattern generation was demonstrated by tuning the ratio of middle- and ring-finger flexion using concurrent two-channel stimulation of their respective forearm motor points. (a) Stimulation amplitudes applied through two channels to subject 1. Five stimulation conditions (A-E) with decreasing channel 1 / channel 2 ratio were tested using the same type of burst kHz waveform (50 Hz burst frequency, 5 ms burst duration, 10 kHz carrier sine wave, Hann window). (b) Simultaneously recorded middle-finger (blue) and ring-finger (green) flexion forces for subject 1 during stimulation conditions A-E. Traces were smoothed, with motion artefacts and baseline offsets removed. (c) Summarised force responses for three subjects, with A-E stimulation amplitudes customised according to each subject's thresholds as described in Section~\ref{sec:ratio}. Each data point represents the mean force during the corresponding stimulation session, while the upper and lower boundaries of the shaded regions indicate the maximum and minimum forces recorded during that session, respectively. Clear ascending ring-finger and descending middle-finger force trends were observed in Subjects 1 and 2. For Subject 3, the ring-finger force remained relatively constant because the motor and maximum tolerable sensory thresholds were close, resulting in only small differences among the Channel 2 amplitudes used in conditions A–E.}
\label{fig8}
\end{figure*}

FFT-based spectral analysis in Fig.~\ref{fig6} shows that the output waveform is indeed dominated by the fundamental component under all tested conditions, with secondary peaks at harmonic frequencies. The fundamental and first three harmonics are marked in red, and an approximate broadband spectral floor is also indicated in dashed lines. Across the tested range, the proportion of harmonic content increases more noticeably with frequency than with amplitude alone. In contrast, the broadband spectral floor varies less markedly across amplitudes, indicating that the residual noise is largely independent upon the output amplitude level.

The spectral quality of the output current was further quantified using THD and SNR, as summarised in Table~\ref{tab:thd_snr}. Across the full parameter range (1-50 mA and 1-50 kHz), THD ranged between 1.2 \% and 15.4 \%, and SNR ranged between 10.8 dB and 38.1 dB. SNR showed a clear increase with amplitude as expected and showed no clear correlation with frequency. At the most common operating conditions (20-50 mA and 1-20 kHz), THD remained below 10 \% and was lower than 2 \% at the lowest frequency. Within the highlighted parameter range, the highest THD occurred at 20 mA and 50 kHz. Overall, waveform quality at low amplitudes was primarily limited by the residual noise floor, whereas distortion became the main source of degradation at high frequencies. Nevertheless, within the highlighted most common operating conditions, the waveform remained well-preserved in terms of both noise and distortion.

The moderate increase in THD at higher carrier frequencies may reflect the finite bandwidth and frequency-dependent behaviour of the closed-loop output stage \cite{vaidyanathan2003theory}. As frequency increases, device parasitics and increased transistor-drive requirements may reduce the ability of the feedback loop to correct waveform errors, resulting in greater residual distortion. Nevertheless, adequate waveform fidelity was maintained throughout the intended stimulation-frequency range.

To assess load dependence, output current waveforms were also measured across five representative resistive and resistive-capacitive (RC) loads used to approximate typical bipolar electrode-tissue interface conditions \cite{vargas2015dynamic}. As shown in Fig.~\ref{fig7}, at 20 kHz the waveform shape is largely preserved across the $1~k\Omega$, $2~k\Omega$ , and RC loads, with visible distortion appearing only at the highest impedance load $10~k\Omega$, where the required output voltage approaches the maximum compliance. At 50 kHz (maximum frequency), distortion is present across all loads, consistent with the high-frequency limitation discussed previously. Within the intended impedance, amplitude, and frequency ranges, SineStim delivers well-preserved waveform shapes for both resistive and RC loads.

\begin{figure}[!t]
\centerline{\includegraphics[width=\columnwidth]{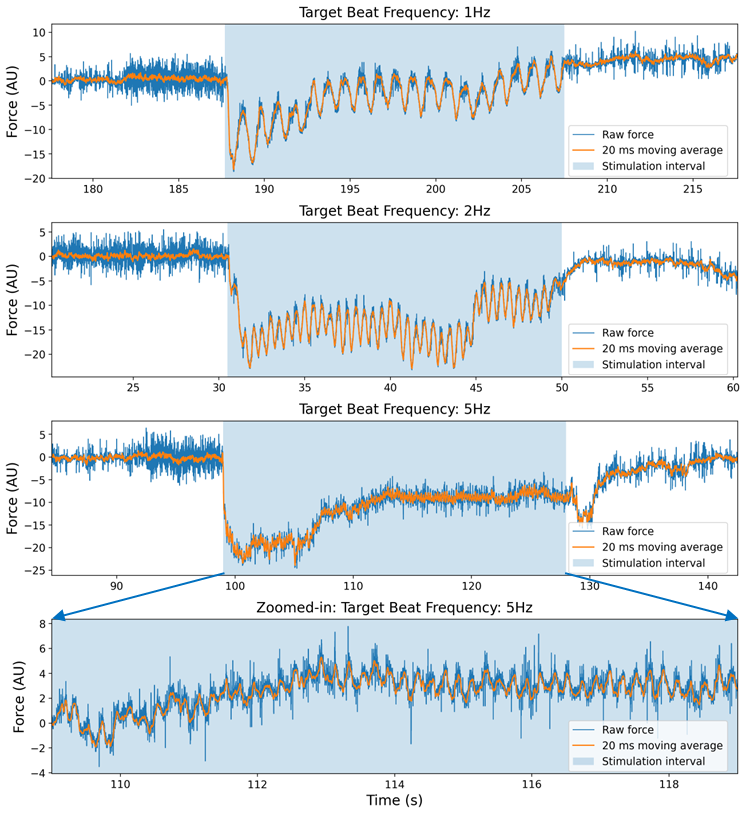}}
\caption{Thumb flexion force recorded during two-channel temporal interference stimulation at the wrist with beat frequencies of 1, 2, and 5 Hz, at 16 mA per channel. Channel 1 was fixed at 10,000 Hz, whereas Channel 2 was set to 10,001, 10,002, or 10,005 Hz to generate the corresponding beat frequencies. Blue traces show the raw force signal in arbitrary units (AU), and orange traces show the 20 ms moving average. Shaded regions denote stimulation-on intervals. Force oscillations at the target beat frequency are clearly visible. At 5 Hz, the modulation depth is reduced, consistent with attenuation by the low-pass mechanical dynamics of the finger, but beat-related modulation remains visible in the zoomed-in view (bottom panel).}
\label{fig9}
\end{figure}

\begin{figure*}[!t]
\centerline{\includegraphics[width=\linewidth]{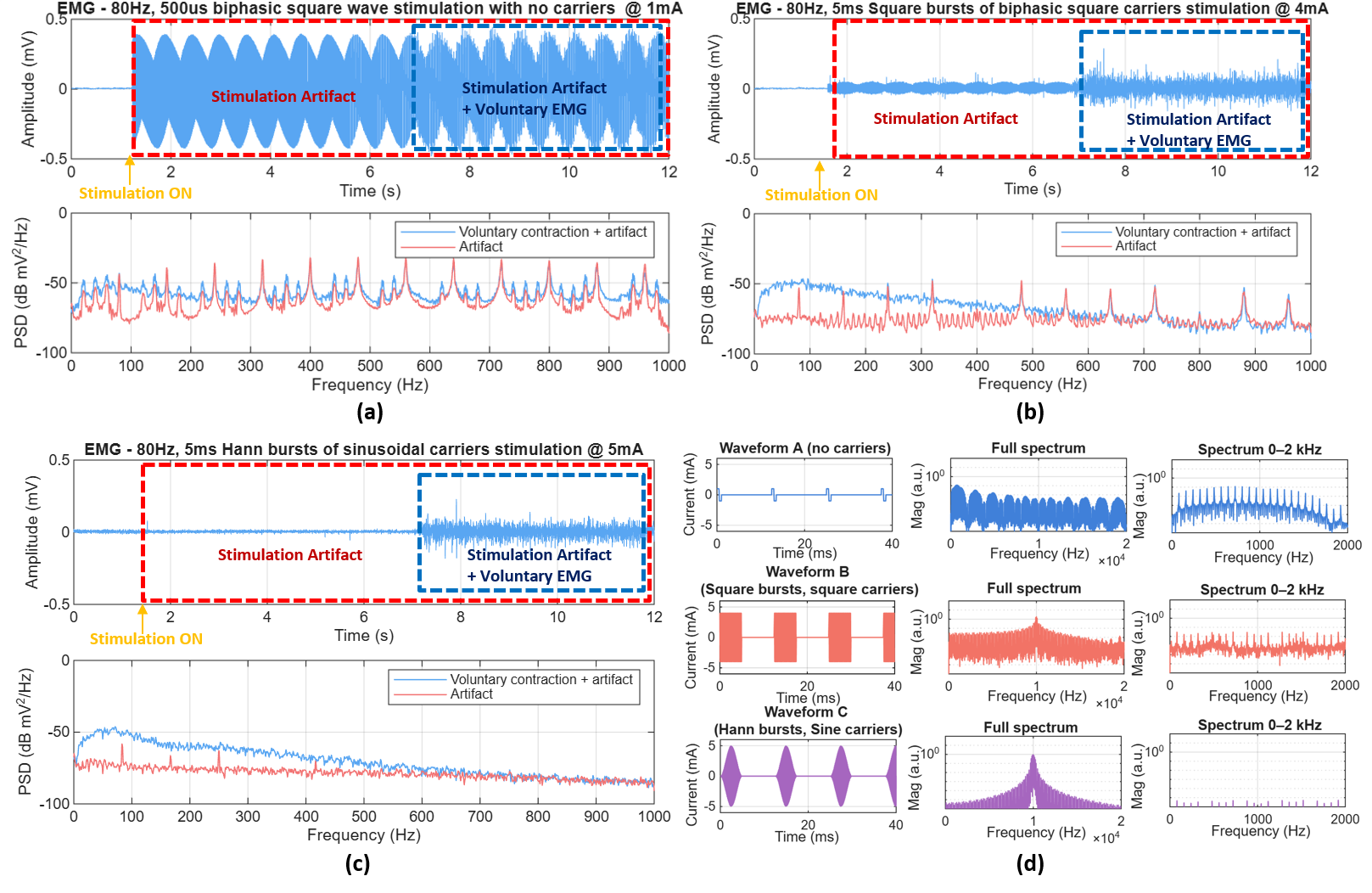}}
\caption{Measured stimulation artifact in concurrent EMG recording during three types of input waveforms delivered at one third of their their respective motor thresholds, with Hann-windowed sine-carrier waveform giving the cleanest EMG recording with the lowest in-band artifact among the three. (a) EMG recording for waveform A and its PSD: conventional biphasic square waveform with no carriers, with 80 Hz frequency and 500 $\mu$s pulse width. Motor threshold is observed at 3 mA and amplitude is set to be 1 mA. Stimulation is applied from the time point marked in yellow, whereas a voluntary contraction of the forearm flexor starts at around 7s. Frequency spectra of the artifact-only region and the artifact + voluntary EMG region are plotted accordingly after being notch filtered at 50Hz. For biphasic square waves, the stimulation artifact completely masks voluntary EMG signals, with frequency components at multiples of the fundamental frequency 80Hz; (b) EMG recording for waveform B and its PSD: square-windowed 10 kHz square-carrier waveform. Motor threshold is observed at 12-13 mA and amplitude is set to be 4 mA. For square burst waves, the artifact is smaller than the previous case but is still visible and comparable to the voluntary EMG amplitude. (c) EMG recording for waveform C and its PSD: Hann-windowed 10 kHz sine-carrier waveform. Motor threshold is observed at around 15 mA and amplitude is set to be 5 mA. In this case, stimulation artifact is substantially smaller than either (a) or (b), with the magnitude much smaller than voluntary EMG level and comparable to background noise. Note that for Hann-windowed sine-carrier stimulation in (c), unlike during steady-state stimulation, transient artifacts sometimes remained at stimulation initiation and termination.
(d) Aforementioned input current waveforms A, B, and C and their corresponding single-sided FFT magnitude spectra. For each condition, the time-domain current waveform, the full FFT magnitude spectrum from 0 to 20 kHz, and the corresponding low-frequency spectral content from 0 to 2 kHz within the EMG recording bandwidth are plotted. The comparison illustrates that square gating and square-wave carriers introduce broader low-frequency and harmonic components, whereas Hann-windowed envelopes and sinusoidal carriers concentrate spectral energy more narrowly around the carrier frequency. At similar motor recruitment levels, the Hann-windowed sine-carrier bursts showed the lowest spectral magnitude within the EMG bandwidth among the tested waveforms. This agrees with the experimental findings in (a) (b) and (c).}
\label{fig10}
\end{figure*}

\subsection{Composite Force Generation}

As shown in Fig.~\ref{fig8}, two-channel motor-point stimulation enabled graded control of the relative flexion forces of the middle and ring fingers. Five stimulation conditions (A-E) were tested using the same burst kHz waveform for all conditions: 50 Hz burst frequency, 5 ms burst duration, 10 kHz sinusoidal carrier, and Hann-windowed envelope. The stimulation amplitudes were customised for each subject according to the motor threshold and maximum tolerable current measured at each motor point. Across conditions A-E, the Channel 1 amplitude, targeting middle-finger flexion, was progressively decreased, while the Channel 2 amplitude, targeting ring-finger flexion, was progressively increased. The amplitude values used for Subject 1 are shown in Fig.~\ref{fig8}(a).

For Subject 1, the simultaneously recorded force traces showed a progressive decrease in middle-finger flexion force and a corresponding increase in ring-finger flexion force across the five stimulation conditions (Fig.~\ref{fig8}(b)). This reciprocal change in force output was also observed in Subject 2, as summarised in Fig.~\ref{fig8}(c). In contrast, Subject 3 showed a relatively constant ring-finger force across conditions. This was consistent with the narrow available stimulation range for the ring-finger motor point in this subject, where the motor threshold and maximum tolerable current were close, resulting in only small differences among the Channel 2 amplitudes used across conditions A--E.

Overall, these results demonstrate that SineStim's waveform is effective at motor recruitment, and is capable of grading relative motor recruitment through multichannel independent amplitude control. This enabled tunable composite motor patterns to be generated from a single multichannel stimulation setup.

\subsection{Temporal Interference Stimulation}
Representative response to TI stimulation at the wrist from one subject is shown in Fig.~\ref{fig9}, where the measured thumb force profile exhibited modulation that tracked the imposed beat frequency. For the 1 Hz and 2 Hz conditions, periodic force fluctuations at the corresponding frequencies were clearly visible during the stimulation intervals, with temporal patterns consistent with the expected beat envelopes. In the 5 Hz condition, beat-related modulation remained present but was attenuated by the mechanical low-pass characteristics of the finger rather than being absent. 

These measurements provide proof-of-concept evidence that SineStim can generate frequency-precise two-channel kilohertz stimulation patterns that translate into observable beat-frequency-modulated motor output in a human peripheral stimulation experiment. These results support the feasibility of using SineStim for experiments requiring independently programmable kilohertz multichannel temporal interference paradigms and other sinusoidal currents across multiple channels.

\subsection{Hann-Windowed Kilohertz Sine-Wave Stimulation Minimises Steady-State Recording Artifact}

Conventional biphasic square-wave stimulation is known to produce large in-band artifacts during concurrent EMG recording. An initial waveform-level FFT analysis was therefore used to provide the rationale for selecting the SineStim waveform for artifact minimisation. As shown in Fig.~\ref{fig10} (d), conventional biphasic square pulses exhibited the largest spectral magnitude within the 0-2~kHz band. For high-frequency carrier bursts, the low-frequency spectral content depended strongly on both the burst envelope and the carrier waveform. Abrupt rectangular gating introduced low-frequency components associated with sharp burst initiation and termination, while square-wave carriers introduced additional harmonic content. These effects were reduced by using a Hann-windowed envelope and a sinusoidal carrier. Accordingly, the Hann-windowed kilohertz sine-carrier burst showed the lowest spectral magnitude within the 0-2~kHz band among the compared input waveforms.

The stimulation artifacts of the three waveforms were then evaluated experimentally with concurrent EMG recordings, as shown in Fig.~\ref{fig10} (a) (b) (c). To compare the artifacts at the same motor recruitment capabilities, three waveforms are compared at one third of their respective motor thresholds, rather than at the same amplitude. As a result, conventional biphasic square-pulse stimulation produced large periodic artifacts, with spectral components at multiples of the 80 Hz stimulation frequency. These artifacts masked the underlying voluntary EMG activity completely in the recorded traces. In contrast, square bursts of 10kHz biphasic square carriers showed reduced artifact, although the artifact level was still comparable to the voluntary EMG. Hann-windowed sine-carrier burst stimulation produced substantially artifact during steady-state stimulation and is the cleanest recording of all three. The frequency-domain analysis in Fig.~\ref{fig10} (a) (b) (c) further supported these observation.

\section{Discussion}
The benchtop characterization shows that SineStim performs best in the typical operating region most relevant and commonly used for motor activation in PES and tSCS, particularly around 1-20 kHz, and above 10 mA. In this range, the waveform remains close to sinusoidal, THD remains moderate, and SNR is high.  Across the full operating range, two limiting behaviours are evident. At low amplitudes below 1 mA, performance is mainly limited by the residual noise floor. At higher frequencies, especially in the range of 30-50 kHz, harmonic distortion limits performance.

The use of a 200 kHz analysis bandwidth for the FFT, THD, and SNR metrics is also justified. The raw oscilloscope time-domain measurements include additional high-frequency oscillation and probe-related artefacts that are not representative of normal stimulation use, since the oscilloscope is not connected during human experiments. In addition, frequency components far above the carrier are expected to be attenuated by the low-pass properties of the electrode-tissue interface and biological tissue \cite{petrofsky2008effect}. For this reason, the band-limited spectral metrics in Fig.~\ref{fig6} are likely more representative of the physiologically relevant delivered waveform than the raw high-bandwidth time traces in Fig.~\ref{fig5} alone. In addition, the measured DC current offset was below 1 \textmu A with no stimulation.

The load-comparison results further indicate that waveform fidelity is preserved across representative resistive and RC loads, with the main exception being the highest resistive load, where the required output voltage approaches the compliance boundary and visible near-saturation distortion appears. This degradation is attributable to compliance-limited voltage swing rather than instability of the current source itself. In practice, higher electrode-skin impedance is associated with smaller electrode area, which in turn requires lower stimulation current to maintain safe current density, thus the compliance range is expected to be sufficient for the majority of practical electrode configurations.


\begin{table*}[!t]
\caption{Comparison of SineStim with similar transcutaneous electrical stimulators for PES and tSCS reported in research literature or available commercially.}
\label{tab:comparison}
\centering
\footnotesize
\setlength{\tabcolsep}{9pt}
\renewcommand{\arraystretch}{1.4}
\begin{tabular}{|>{\raggedright\arraybackslash}m{1.8cm}|>{\centering\arraybackslash}m{1.45cm}|>{\centering\arraybackslash}m{1.25cm}|>{\centering\arraybackslash}m{1.2cm}|>{\centering\arraybackslash}m{1.4cm}|>{\centering\arraybackslash}m{1.4cm}|>{\centering\arraybackslash}m{1.45cm}|>{\centering\arraybackslash}m{1.4cm}|>{\centering\arraybackslash}m{1.4cm}|}
\hline
\textbf{Feature} &
\makecell{\textbf{SineStim} \\(Research \\Grade)} &
\makecell{\textbf{PulseStim}\\ \cite{sheeraz2026bidirectional}\\(Research \\Grade)} &
\makecell{\textbf{Custom}\\\textbf{Stimulator} \\ \cite{trout2023portable}\\(Research \\Grade)} &
\makecell{\textbf{OpenXstim} \\ \cite{alam2025openxstim}\\(Research \\Grade)} &
\makecell{\textbf{NeoStim-C.5$^{*}$} \\ \cite{grishin2017five}\\\textbf{(Cosyma)}\\(Commercial)} &
\makecell{\textbf{STG5}\\\textbf{(Multichannel} \\\textbf{Systems)}\\(Commercial, \\not for human)
} &
\makecell{\textbf{DS8R}\\\textbf{(Digitimer)}\\ (Commercial)} &
\makecell{\textbf{DS5} \\\textbf{(Digitimer)}\\ (Commercial)} \\
\hline
Channel count & 12 & 4 & 3 & 2 & 5 & 2 & 1 & 1 \\
\hline
Voltage compliance & $\pm$120 V & 72 V & $\pm$150 V & 96 V & -- & 70 V & 400 V & $\pm$120 V \\
\hline
Waveform &
\makecell{Custom \\bursts w/ \\kHz sinusoidal \\carriers}&
\makecell{Square bursts \\w/o carriers} &
\makecell{Square bursts\\w/o carriers} &
\makecell{Square bursts \\w/ kHz square \\carrier} &
\makecell{Square bursts \\w/ kHz square \\carrier} &
\makecell{Rectangular,\\ramp, or\\sinusoidal} &
\makecell{Square bursts \\w/ kHz square \\carrier} &
\makecell{Defined by\\external\\analogue\\command} \\
\hline
Current-controlled output & Yes & Yes & Yes & Yes & Yes & Yes & Yes & Yes \\
\hline
\makecell[l]{Max amplitude\\and resolution} &
\makecell{50 mA\\$\pm$0.1 mA} &
\makecell{20 mA\\$\pm$0.1 mA} &
\makecell{6 mA\\--} &
\makecell{110 mA\\--} &
\makecell{250mA\\--} &
\makecell{16 mA\\$\pm$0.6 $\mu$A} &
\makecell{1000 mA\\$\pm$0.1 mA} &
\makecell{50 mA\\trigger source-\\dependent} \\
\hline
\makecell[l]{Frequency range\\and resolution} &
\makecell{0--50 kHz\\$\pm$0.1 Hz} &
\makecell{0--250 Hz\\$\pm$1 Hz} &
\makecell{1--200 Hz\\--} &
\makecell{0--9 kHz\\--} &
\makecell{5--10 kHz\\--} &
\makecell{0--100 kHz\\--} &
\makecell{0--10 kHz\\$\pm$1\%} &
\makecell{0--50 kHz\\trigger source-\\dependent} \\
\hline
Dimensions (mm) &
\makecell{280$\times$180\\$\times$140\\(enclosure)} &
\makecell{100$\times$120\\(bare board)} &
\makecell{59$\times$56\\(bare board)} &
\makecell{--} &
\makecell{--} &
\makecell{325$\times$310\\$\times$47\\(enclosure)} &
\makecell{225$\times$100\\$\times$255\\(enclosure)} &
\makecell{225$\times$100\\$\times$255\\(enclosure)} \\
\hline
\makecell[l]{THD \\(sinusoidal)} &
\makecell{1 - 8 \%} &
-- & -- & -- & -- & -- & -- & -- \\
\hline
\makecell[l]{External analogue \\ driver device \\required} &
No & No & No & No & No & No & Yes & Yes \\
\hline
Real-time current monitoring & Yes & Yes & No & Yes & -- & No & Yes & Yes \\
\hline
\multicolumn{9}{|>{\raggedright\arraybackslash}p{\textwidth}|}{\footnotesize $^{*}$ BioStim-5 in earlier literature.}\\
\hline
\end{tabular}
\end{table*}

The two-channel human subject stimulation experiments extended the benchtop characterization by showing that the stimulator can generate two-channel kilohertz waveforms with both independently-tuned amplitude ratio and frequency offset that produce distinct motor output patterns in human participants. The capacities naturally extend to more channels. If combined with finite element modelling, multichannel parameter optimisation for spatial stimulation targeting and simultaneous stimulation control of multiple pathways are made possible. With artifact-minimised EMG recordings, closed-loop stimulation control based on real-time EMG measurements can be implemented without loss of EMG information due to blanking. For single-pulse stimulation, Hann-windowed sine carrier stimulation may not be advantageous compared to single biphasic square wave, because the latter is clean for separating evoked EMG in offline analysis. However, for potential closed-loop and real-time applications where sustained stimulation is needed for training and rehabilitation, continuous stimulation with minimal steady-state recording artifact provides a stable and simple solution, without relying on synchronisation of triggering signals or separation of EMG and stimulation intervals. Furthermore, although the performance is demonstrated via PES here, the combination of compliance, frequency range, and multichannel programmability naturally suggests applicability to tSCS applications which can be evaluated in the future.

Comparison with commercially available stimulators and similar research-grade devices reported in the literature is summarized in Table~\ref{tab:comparison}. Among the compared systems, SineStim provides the highest channel count. Among the few devices capable of generating pure sinusoidal waveforms, namely SineStim, STG5, and DS5, SineStim combines a wide voltage compliance, current output range, and frequency range with multichannel capability, portability, without requiring an external signal source for waveform generation. Compared with the commonly-used commercial Digitimer stimulators DS8R and DS5, SineStim matches DS5 in the output range most relevant to kilohertz transcutaneous stimulation, while additionally offering substantially higher channel count, portability, and direct PC control. By contrast, the custom multichannel stimulator reported in \cite{trout2023portable} and our previously developed PulseStim platform \cite{sheeraz2026bidirectional} only support square-pulse stimulation without carrier modulation. Compared with OpenXstim \cite{alam2025openxstim} and NeoStim-C.5 \cite{grishin2017five} which both can deliver square bursts with kHz square carriers, SineStim offers a wider frequency range, has a higher channel count, and enables pure sinusoidal carriers which are advantageous for reducing in-band EMG artifact. Overall, these characteristics make SineStim a unique platform for future studies on multichannel stimulation optimisation, high-frequency stimulation, and closed-loop EMG-based stimulation control.

\section{Conclusion}

SineStim, a portable, 12-channel, kilohertz-frequency current-controlled neurostimulator for non-invasive peripheral nerve stimulation, with a novel architecture, ±120 V compliance and per-channel isolation was designed and successfully validated. Benchtop characterisation demonstrated a total harmonic distortion ranging from 1 \% to 8 \% and confirmed a signal-to-noise ratio exceeding 38 dB for typical operating conditions. Human subject tests further demonstrated the device's capacity for multichannel frequency and amplitude tuning and its effectiveness in eliciting controlled motor response. Minimal-artifact stimulation offers another compelling reason for using kilohertz sinusoidal stimulation strategies over conventional waveforms. SineStim provides a novel multichannel noninvasive stimulation platform that has the potential to facilitate future studies in spatio-temporal stimulation optimisation and EMG-controlled closed-loop neurostimulation.

\section*{Acknowledgment}
We would like to thank Dr Agnese Grison for providing the finger force platform.

\bibliographystyle{IEEEtran}


\end{document}